\documentclass[12pt]{article}
\newcommand{\beq}{\begin{equation}}
\newcommand{\eeq}{\end{equation}}
\begin{document}

\vspace{1cm}

\begin{center}
{\bf ON THE THEORY OF NUCLEAR ANAPOLE MOMENTS}
\end{center}

\begin{center}
V.F. Dmitriev\footnote{e-mail address: dmitriev@inp.nsk.su}, and
I.B. Khriplovich\footnote{e-mail address: khriplovich@inp.nsk.su}
\end{center}
\begin{center}
Budker Institute of Nuclear Physics, 630090 Novosibirsk,
Russia
\end{center}

\bigskip

\begin{abstract}
We discuss the present state of the theory of nuclear anapole
moments.
\end{abstract}

Forty-five years ago it was pointed out \cite{zel} that in a
system which has no definite parity a special distribution of
magnetic field may arise. It cannot be reduced to common
electromagnetic multipoles, such as dipole or quadrupole moments,
but looks like the magnetic field created by a current in toroidal
winding. This special source of electromagnetic field  was called
(by A.S. Kompaneets) ``anapole''.

For many years the anapole remained a theoretical curiosity only.
The situation has changed due to the investigation of parity
nonconservation (PNC) in atoms. Since these tiny effects increase
with the nuclear charge $Z$, all the experiments are carried out
with heavy atoms. The main contribution to the effect is
independent of nuclear spin and caused by the parity-violating
weak interaction of electron and nucleon neutral currents.  This
interaction is proportional to the so-called weak nuclear charge
$Q$ which is numerically close (up to the sign) to the neutron
number $N$. Thus, in heavy atoms the nuclear-spin-independent
weak interaction is additionally enhanced by about two orders of
magnitude. Meanwhile, the nuclear-spin-dependent effects due to
neutral currents not only lack the mentioned coherent
enhancement, but are also strongly suppressed numerically in the
electroweak theory. Therefore, the observation of PNC
nuclear-spin-dependent effects in atoms looked absolutely
unrealistic.

However, it was demonstrated \cite{fk,fks} that these effects in
atoms are dominated not by the weak interaction of neutral
currents, but by the electromagnetic interaction of atomic
electrons with nuclear anapole moment (AM). It should be mentioned
first of all that the magnetic field of an anapole is contained
within it, in the same way as the magnetic field of a toroidal
winding is completely confined inside the winding. It means that
the electromagnetic interaction of an electron with the nuclear AM
occurs only as long as the electron wave function penetrates the
nucleus. In other words, this electromagnetic interaction is as
local as the weak interaction, and they cannot be distinguished by
this interaction. The nuclear AM is induced by PNC nuclear forces
and is therefore proportional to the same Fermi constant
$G=1.027\times 10^{-5} m^{-2}$ (we use the units $\hbar=1, c=1$;
$m$ is the proton mass), which determines the magnitude of the
weak interactions in general and that of neutral currents in
particular. The electron interaction with the AM, being of the
electromagnetic nature, introduces an extra small factor into the
effect discussed, the fine-structure constant $\alpha=1/137$.
Then, how it comes that this effect is dominating? The answer
follows from the same picture of a toroidal winding. It is only
natural that the interaction discussed is proportional to the
magnetic flux through such a winding, and hence in our case to the
cross-section of the nucleus, i.e. to $A^{2/3}$ where $A$ is the
atomic number. In heavy nuclei this enhancement factor is close to
30 and compensates essentially for the smallness of the
fine-structure constant $\alpha$. As a result, the dimensionless
effective constant $\kappa$ which characterizes the anapole
interaction in the units of $G$ is not so small in heavy atoms,
but is numerically close to 0.3 (we use the same definition of the
effective constant $\kappa$ as in \cite{fk,fks}).

The nuclear anapole moment was experimentally discovered in 1997
\cite{woo}. This result for the total effective constant of the
PNC nuclear-spin-dependent interaction in $^{137}$Cs is
\beq
\kappa_{tot}=0.44(6).
\eeq
To extract this number from experimental data, the results of
atomic calculations \cite{fra,kra} were used; these calculations
performed using different approaches are in excellent agreement,
and there are good reasons to believe that their accuracy is no
worth than 2-3\%. If one excludes the neutral current
nuclear-spin-dependent contribution from the above number, as
well as the result of the combined action of the ``weak'' charge
$Q$ and the usual hyperfine interaction, the answer for the
anapole constant will be
\beq\label{exp}
\kappa=0.37(6).
\eeq
Thus, the existence of an AM of the $^{137}$Cs nucleus is
reliably established. A beautiful new physical phenomenon, a
peculiar electromagnetic multipole has been discovered.

But the discussed result does not reduce to only this. It brings
valuable information on PNC nuclear forces. Of course, to this end
it should be combined with reliable nuclear calculations. However,
it is instructive to start their discussion, as it was done in
\cite{fks}), with a rather crude approximation. Not only one
assumes here that the nuclear spin $\mathbf{I}$ coincides with the
total angular momentum of an odd valence nucleon, while the other
nucleons form a core with the zero angular momentum. The next
assumption is that the core density $\rho(r)$ is constant
throughout the space and coincides  with the mean nuclear density
$\rho_0$. The last assumption, ascending to \cite{cur}, is
reasonable if the wave function of the external nucleon is mainly
localized in the region of the core. Then simple calculations give
the following result for the anapole constant \cite{fks}:
\begin{equation}\label{an}
\kappa=\frac{9}{10}\,g\,\frac{\alpha \mu}{m r_0}\, A^{2/3}.
\end{equation}
Here $g$ is the effective constant of the P-odd interaction of
the outer nucleon with the nuclear core, $\mu$ is the magnetic
moment of the outer nucleon, $r_0=1.2$ fm. The $A$-dependence of
this constant is very natural. Indeed, since the anapole
corresponds to the magnetic field configuration induced by a
toroidal winding, the AM value should be proportional to the
magnetic flux, i.e., to the cross-section area of the nucleus,
hence to $A^{2/3}$.

The so-called ``best values'' for the parameters of P-odd nuclear
forces \cite{ddh} result in $g_p=4.5$ for an outer proton
\cite{fks,dfst,fts}. Thus obtained values for AMs of the nuclei of
experimental interest are presented in the first column of Table.

\begin{table}
[h]
\begin{center}
\begin{tabular}{|c|c|c|c|c|c|c|c|c|c|c|} \hline

&&&&&&&&&&\\
   & \cite{fks}$^{a)}$ & \cite{fks}$^{b)}$& \cite{hhm}$^{c)}$&\cite{bp,bp1}$^{d)}$
&\cite{dkt}$^{e)}$&\cite{dkt}$^{f)}$&\cite{dt}$^{g)}$&\cite{ab}&\cite{dt1}
&\cite{hm}\\
&&&&&&&&&&\\
\hline
&&&&&&&&&&\\
$^{133}$Cs & 0.37 & 0.28 & 0.14 (0.28) &0.27& 0.33&0.26&0.22&---&0.15&0.21\\
&&&&&&&&&&\\
\hline &&&&&&&&&&\\
$^{203,205}$Tl &0.49& 0.43 &---&0.27&0.42&0.40&0.37&0.24&0.24&0.10\\
&&&&&&&&&&\\
\hline
&&&&&&&&&&\\
$^{209}$Bi &  0.51 & 0.35 & ---&0.27 &0.45&0.29&0.30&---&0.15&---\\
&&&&&&&&&&\\
\hline
\end{tabular}
\end{center}
\begin{center}
{\bf Table}
\end{center}
{\it a)Calculation with formula (\ref{an}). b)Woods-Saxon
potential including spin-orbit interaction. c)Only the P-odd
$\pi$-meson-exchange was calculated in \cite{hhm}; we indicate in
brackets what to our guess would be the result of \cite{hhm} if
the P-odd short-range were included. d)Our extrapolation of the
results of \cite{bp,bp1}, from their values of $g_p$ to $g_p=4.5$.
e)Oscillator potential, contribution of contact current included.
f)Woods-Saxon potential, contributions of contact and spin-orbit
currents included. g)Many-body corrections calculated in the
constant-density approximation.}
\end{table}

Various calculations of the nuclear AMs, going beyond simple
analytical formula (\ref{an}), (see the results in Table) can be
roughly divided into two groups: the calculations within the
independent particle model (IPM) using Woods-Saxon and oscillator
potentials [3, 12-14], and the calculations including many-body
effects [11, 15-18]. In fact, some of the many-body contributions
were discussed in \cite{bp,bp1} as well.

The analytical estimate (\ref{an}) produces smooth $A^{2/3}$
behaviour, but certainly exaggerates the effect due to the
assumption that the P-odd contact interaction with the nuclear
core extends throughout the whole localization region of the
unpaired nucleon. Indeed, the IPM calculations reveal certain
shell effects quite pronounced in the values of $\kappa$ for Tl
and Bi (see Table). Both these nuclei are close to the
doubly-magic $^{208}$Pb. However, while the anapole moment of Tl
nucleus in IPM is close to its analytical estimate, the anapole
moment of Bi in IPM differs significantly both from the analytical
formula and from the anapole moment of Tl. This difference can be
attributed to the difference in the single-particle orbitals for
the unpaired proton in Tl and Bi. The 3s$_{1/2}$ wave function in
Tl is concentrated essentially inside the nuclear core, while the
1h$_{9/2}$ wave function in Bi is pushed strongly outside of it.
By this reason the unpaired proton in Bi ``feels'' in fact much
smaller part of the P-odd weak potential. An analogous suppression
of the PNC interaction takes place for the outer 1g$_{7/2}$ proton
in Cs.

Various approaches were used as well in the many-body calculations
[11, 15-18]. In one of them \cite{dt,dt1} the random-phase
approximation (RPA) with effective forces was employed to
calculate the effects of the core polarization. In another
approach \cite{hhm,hm} large basis shell-model calculations were
performed. However, in the last case there is a serious problem:
the basis necessary to describe simultaneously the effects of both
regular nuclear forces and P-odd ones is in fact too large.
Therefore, some additional approximations were made in
\cite{hhm,hm} in order to reduce the size of the basis space.

Fortunately, the Tl nucleus is a rather special case in the
many-body approach as well. Not only is it close to the
doubly-magic $^{208}$Pb, but its unpaired proton is 3s$_{1/2}$,
but not 1h$_{9/2}$ as in Bi. This makes the effects of the core
polarization here relatively small. Thus the density of states in
Tl is reduced, and an effective Hamiltonian suitable for
shell-model calculations can be constructed \cite{rblp}. This
Hamiltonian was used in \cite{ab} to calculate the anapole moment
of Tl nucleus. The result of \cite{ab} and the RPA result of
\cite{dt1} for the thallium coincide, in spite of completely
different descriptions of nuclear forces used in these works to
calculate the core polarization. These results of \cite{ab,dt1}
differ essentially from the value obtained in \cite{hm} under
extra assumptions: the closure approximation and further reduction
of a three-body matrix element to the two-body one. It is also
worth mentioning perhaps that in \cite{ab,dt1} and \cite{hm}
different parameterizations of the parity violating nuclear forces
have been used.

Thus we believe that the theoretical predictions for the AMs of
nuclei of the present experimental interest, can be reasonably
summarized now, at ``best values'' of P-odd constants, as follows:
\beq\label{pre}
\kappa(^{133}{\rm Cs})=0.15-0.21, \quad \kappa(^{203,205}{\rm
Tl})=0.24, \quad \kappa(^{209}{\rm Bi})=0.15.
\eeq
We believe also that there are good reasons to consider these
predictions as sufficiently reliable, at the accepted values of
the P-odd nuclear constants.

The comparison of the value (\ref{pre}) for the cesium AM with
the experimental result (\ref{exp}) indicates that the ``best
values'' of \cite{ddh} somewhat underestimate the magnitude of
P-odd nuclear forces. In no way is this conclusion trivial. The
point is that the magnitude of parity-nonconserving effects found
in some nuclear experiments is much smaller than that following
from the ``best values'' (see review \cite{ah}). In all these
experiments, however, either the experimental accuracy is not
high enough, or the theoretical interpretation is not
sufficiently convincing. The experiment \cite{woo} looks much
more reliable in both respects.

Here it should be mentioned however that the experimental result
for the thallium AM, $\kappa =-0.22\pm 0.30$ \cite{vet}, does not
comply with the theoretical prediction for it presented in
(\ref{pre}) (the disagreement will be even more serious if one
assumes that the nuclear P-odd constants are larger than the
``best values'' of \cite{ddh} as indicated by the measurement of
the cesium AM). Obviously, it is highly desirable for this problem
to be cleared up.

Clearly, in line with its general physics interest, the
investigation of nuclear AMs in atomic experiments is first-rate,
almost table-top nuclear physics.

\bigskip

{\bf Acknowledgements} We are grateful to Yu.P. Gangrsky and B.N.
Markov for their interest to the problem and the suggestion to
write this note. The work was supported in part by the Grant No.
00-15-96811 for Leading Scientific Schools, and by the Federal
Program Integration-2001.


\begin{thebibliography}{99}

\bibitem{zel} Ya.B. Zel'dovich, Zh. Eksp. Teor. Fiz. {\bf 33}
(1957) 1531 [Sov. Phys. JETP {\bf 6} (1957) 1184] (the paper
contains also the mention of the analogous results obtained by
V.G. Vaks.)
\bibitem{fk} V.V. Flambaum, I.B. Khriplovich, Zh. Eksp. Teor. Fiz. {\bf 79}
(1980) 1656\\ $[$Sov. Phys. JETP {\bf 52} (1980) 835]
\bibitem{fks} V.V. Flambaum, I.B. Khriplovich, O.P. Sushkov, Phys. Lett.
{\bf B146} (1984) 367
\bibitem{woo} C.S. Woods et al., Science {\bf 275} (1997) 1759
\bibitem{fra} P.A. Frantsuzov, I.B. Khriplovich, Z. Phys. {\bf D7} (1988) 297
\bibitem{kra} A.Ya. Kraftmakher, Phys. Lett. {\bf A132} (1988) 167
\bibitem{cur} F. Curtis Michel, Phys. Rev. {\bf B133} (1964) 329
\bibitem{ddh} B. Desplanques, J.F. Donoghue, B.R. Holstein, Ann. Phys.
{\bf 124} (1980) 449
\bibitem{dfst} V.F. Dmitriev et al., Phys. Lett. {\bf B125} (1983) 1
\bibitem{fts} V.V. Flambaum, V.B. Telitsin, O.P. Sushkov, Nucl. Phys.
{\bf A444} (1985) 611
\bibitem{hhm} W.C. Haxton, E.M. Henley, M.J. Musolf, Phys. Rev. Lett.
{\bf 63} (1989) 949
\bibitem{bp} C. Bouchiat, C.A. Piketty, Z. Phys. {\bf C49} (1991) 91
\bibitem{bp1} C. Bouchiat, C.A. Piketty, Phys. Lett. {\bf B269} (1991) 195;
erratum {\bf B274} (1992) 526
\bibitem{dkt} V.F. Dmitriev, I.B. Khriplovich, V.B. Telitsin, Nucl. Phys.
{\bf A577} (1994) 691
\bibitem{dt} V.F. Dmitriev, V.B. Telitsin, Nucl. Phys. {\bf A613} (1997) 237
\bibitem{ab} N. Auerbach, B.A. Brown, Phys. Rev. {\bf C60} (1999) 025501
\bibitem{dt1} V.F. Dmitriev, V.B. Telitsin, Nucl. Phys. {\bf A674} (2000)
168
\bibitem{hm} W.C. Haxton, C.-P. Liu, M.J. Ramsey-Musolf, nucl-th/0109014
\bibitem{rblp} L. Rydstrom et al., Nucl. Phys. {\bf A512} (1990) 217
\bibitem{ah} E.G. Adelberger, W.C. Haxton, Ann.Rev.Nucl.Part.Sci. {\bf 35}
(1985) 501
\bibitem{vet} P.A. Vetter et al., Phys. Rev. Lett. {\bf 74} (1995) 2658

\end{thebibliography}
\end{document}